\documentclass[10pt,conference]{IEEEtran}

\usepackage{etex} \reserveinserts{28}

\usepackage[english]{babel}
\usepackage{xcolor}
\usepackage{hyphenat}

\newcommand\myworries[1]{}
\newcommand\comments[1]{}
\newcommand\dg[1]{}
\newcommand\vb[1]{}

\usepackage{listings}
\usepackage{color}
\usepackage{latexsym}
\usepackage{mathtools}

\usepackage{amsthm}

\usepackage{amsfonts}
\usepackage{graphicx}
\usepackage{flushend}
\usepackage{siunitx}
\usepackage{bigfoot}
\usepackage{hyperref}

\usepackage{wrapfig}

\usepackage{enumitem}

\usepackage[linesnumbered,noend]{algorithm2e}

\usepackage{caption}

\usepackage{epstopdf}

\usepackage[export]{adjustbox}

\usepackage{amssymb}
\usepackage{pifont}
\newcommand{\xmark}{ }%

\usepackage{cleveref}

\definecolor{dkgreen}{rgb}{0,0.6,0}
\definecolor{gray}{rgb}{0.5,0.5,0.5}
\definecolor{mauve}{rgb}{0.58,0,0.82}

\usepackage{tikz-cd}
\usetikzlibrary{decorations.pathmorphing}

\newcommand\xrsquigarrow[1]{%
    \mathrel{%
        \begin{tikzpicture}[%
            baseline={(current bounding box.south)}
            ]
        \node[%
            ,inner sep=.44ex
            ,align=center
            ] (tmp) {$\scriptstyle #1$};
        \path[%
            ,draw,<-
            ,decorate,decoration={%
                ,zigzag
                ,amplitude=0.7pt
                ,segment length=1.2mm,pre length=3.5pt
                }
            ] 
        (tmp.south east) -- (tmp.south west);
        \end{tikzpicture}
        }
    }

\newcommand\xqed[1]{%
  \leavevmode\unskip\penalty9999 \hbox{}\nobreak\hfill
  \quad\hbox{#1}}
\newcommand\closeExample{\xqed{$\triangle$}}

\title{Model Checker Execution Reports}
\author{
Rodrigo Casta\~no 
\qquad
Victor Braberman
\qquad
Diego Garbervetsky
\qquad
Sebastian Uchitel\\
Departamento de Computaci\'on,\\
FCEyN, UBA, CONICET\\
Buenos Aires, Argentina\\
\{rcastano,diegog,vbraber,suchitel\}@dc.uba.ar
}

\newtheorem{mydef}{Definition}
\newtheorem{myprop}{Property}
\newtheorem{example}{Example}

\begin{document}

\maketitle
\begin{abstract}
Software model checking constitutes an undecidable problem and, as such, even an ideal tool will in some cases fail to give a conclusive answer. In practice, software model checkers fail often and usually do not provide any information on what was effectively checked. 
The purpose of this work is to provide a conceptual framing to extend software model checkers in a way that allows users to access information about incomplete checks.
We characterize the information that model checkers themselves can provide, in terms of analyzed traces, i.e.\ sequences of statements, and \emph{safe cones}, and present the notion of execution reports, which we also formalize.
We instantiate these concepts for a family of techniques based on Abstract Reachability Trees and implement the approach using the software model checker CPAchecker.
We evaluate our approach empirically and provide examples to illustrate the execution reports produced and the information that can be extracted.

\end{abstract}

%

\hyphenation{key-span CPA-check-er}

\section{Introduction}
\label{sec:Intro}
Software model checking~\cite{jhala2009software} constitutes an undecidable problem and, as such, even an ideal tool will in some cases fail to give a conclusive answer.
In practice, undecidability is not the only issue.
The vast state spaces can lead to the complete exhaustion of system resources or impractically long execution times.

Software model checking has been making steady progress during the past decade and today's state-of-the-art software model checkers can handle specific industrial problems particularly well.
For instance SDV~\cite{ball2011decade} is highly successful in finding bugs in Windows device drivers.

Unfortunately, some instances take hours of computation, only to inform the user that no counterexample was found within the allotted time or memory limit.
A user facing this situation is confronted with several high-level questions about what the verification attempt actually achieved.
Should she retry with a longer time limit? How much longer? Is the  tool making any progress on the instance? Maybe she should try another technique?

Our goal is to extend and complement existing work on partial verification by providing a different way for users to observe the work performed by the software model checker.
An important step towards our goal is to be able to answer much simpler inquiries about incomplete verification attempts.

We believe answering  the following informal questions would be valuable for a user after an inconclusive verification attempt:
\begin{itemize}
\item Can partial safety assurances about the system be extracted from an incomplete verification attempt?
For instance, a user that receives a report showing that a whole class of relevant behaviors has been exhaustively checked may use this as part of a dependability case. 
\item Can behavior that was not analyzed be explored by a user?
	For instance, a user that can observe that relevant classes of behavior have not even been looked at by the checker, let alone verified, may decide that what seemed like a sufficiently thorough analysis is not such (e.g., an inexperienced user would benefit from knowing that a tool based on BMC with a fixed bound can sometimes give up without ever exploring anything beyond an initialization loop\cite{lal2014powering}). Moreover, a more experienced user attempting full verification may decide that a drastic change in the verification strategy is needed (e.g., another tool, abstracting the system\hyp under\hyp verification, etc.).
\end{itemize}

By incomplete verification attempt we refer to a situation when a software model checker fails to confirm any counterexample as feasible and also fails to prove the instance safe.

In this paper we explore answering the first question using the notion of \emph{safe cone}. A \emph{safe cone} is a finite trace for which any extension has been analyzed in the incomplete verification attempt. A \emph{minimal feasible} \emph{safe cone} represents compactly a set of traces that have been successfully verified by the checker. For the second question we build on the notion of a \emph{frontier}. A \emph{frontier trace} is a feasible finite trace that was analyzed by the checker but none of its feasible extensions were. Frontier traces represent compactly classes of traces that were not explored by the checker.

Our hypothesis is that execution reports that under\hyp approximate the set of minimal feasible \emph{safe cones} and the set of maximal feasible \emph{frontier traces} can be computed in reasonable time (with respect to the cost of verification) and can provide non-trivial feedback on incomplete verification attempts.

We will start by illustrating with examples the information we wish to extract and defining a possible formalization of the idealized properties it should have.
Subsequently, we define and discuss Execution Reports, an under-approximation of the ideal output.
Afterwards, we  instantiate these concepts for the family of techniques based on Abstract Reachability Trees (ART)~\cite{henzinger2002lazy}, and discuss a proof-of-concept implementation built as an extension of CPAchecker~\cite{beyer2007configurable}. We also include an empirical evaluation of our implementation.

We conclude this paper with a discussion of related work and how our approach compares to existing techniques followed by a few concluding remarks.


\section{Motivation: What has and has not been analyzed?}
\label{sec:motivation}

We frame our work in the context of a verification attempt being interrupted before its completion.

Many verification techniques work by incrementally exploring the state space.
This is the case for software model checking techniques and implementations like BMC~\cite{biere1999symbolic}, lazy predicate abstraction~\cite{henzinger2002lazy}, inlining and unrolling based-techniques like Corral~\cite{lal2012solver}, DSE~\cite{cadar2013symbolic} and, in general, ART-based implementations of various techniques, including explicit value analysis~\cite{beyer2013explicit} and CEGAR variants of some of the previous among other. 

In these techniques and implementations one can understand that incremental exploration leads to an incremental but silent increase of $analyzed$ traces, i.e. sequences of statements, as depicted in the examples that follow. 
Moreover, certain statement traces will reach a portion of the behavior space that has been fully verified, implicitly defining a \emph{safe cone} containing all possible ways of extending such traces. 

Understanding that partial explorations provide safety assurances, we are particularly interested in the \emph{frontiers} defined by minimal \emph{safe cone} traces and maximal $analyzed$ statement traces.

We illustrate these concepts with a verification attempt of the instance in Figure~\ref{fig:basic_array} with bounded model checking.

\begin{figure}
\begin{lstlisting}
int nondet();
int min(int a[], int n) {
    int res = a[0];
    for (int i=0;i<n;++i)
      if (a[i] < res) res = a[i];
    return res;
}
void init_vector(int a[], int n) {
    %*\label{line:before_loop}*)int i = 0;
    %*\label{line:for_loop}*)for (i=0;i<n;++i) {
      %*\label{line:for_assignment}*)a[i] = nondet();
    }
}
void test_min(int large) {
    %*\label{line:safe_cone_line1}*)int n;
    %*\label{line:safe_cone_line2}*)if (large) {
        %*\label{line:set_n_large}*)n = 20;
    } else {
        %*\label{line:set_n_small}*)n = 1;
    }
    %*\label{line:after_set_n_small1}*)int a[20];
    %*\label{line:after_set_n_small2}*)init_vector(a, n);
    int min_elem = min(a, n); 
    assert(min_elem <= a[0]);
}
\end{lstlisting}
\caption{Harness for method \texttt{min}}
\label{fig:basic_array}
\end{figure}
\begin{example}[Analyzed behaviors in BMC]
    \label{ex:analyzedBMC}

The code snippet in Figure~\ref{fig:basic_array} corresponds to a parametric test harness exercising the method \texttt{min}.
The harness input as well as the result of method \texttt{nondet} are interpreted by the verification technique as non-deterministic.
Our verification attempt, in this example, is not interrupted due to reaching a resource limit but instead due to using a bounded model checker~\cite{biere1999symbolic} with a bound on the number of loop iterations set to 3.

Using this configuration, the tool would perform an exhaustive exploration but would disregard any executions involving the fourth loop iteration of the \texttt{for}-loop in \texttt{init\_vector}.

Any sequence of statements reaching line~\ref*{line:set_n_small} (\texttt{n = 1;}) will necessarily satisfy the safety property, since the incomplete verification attempt would not find any assertion failures and the loop within \texttt{init\_vector} can be exhaustively analyzed.
That means the sequence of statements consisting of lines~\ref*{line:safe_cone_line1} (\texttt{int n;}),~\ref*{line:safe_cone_line2} (\texttt{if (large)}) and~\ref*{line:set_n_small} (\texttt{n = 1;}) defines a \emph{safe cone}, because every continuation of the statement trace is also $safe$.
The former trace is also minimal in the sense that the trace resulting from removing the last statement, line~\ref*{line:set_n_small}, is not a \emph{safe cone}.

	Moreover, any execution that carries out line~\ref*{line:for_loop} (\texttt{for} \texttt{(i=0;} \texttt{i<n;} \texttt{++i)}) at most 4 times (3 full iterations and 1 bound check) is also analyzed by the incomplete verification attempt.
In contrast, any sequence containing line~\ref*{line:for_loop} at least 5 times is ignored by BMC and therefore will not be examined.
Therefore, $analyzed$ does not contain traces exercising line~\ref*{line:for_loop} 5 times or more.
That is, the trace composed of lines~\ref*{line:safe_cone_line1},~\ref*{line:safe_cone_line2},~\ref*{line:set_n_large},~\ref*{line:after_set_n_small1},~\ref*{line:after_set_n_small2},~\ref*{line:before_loop} and then 4 repetitions of lines~\ref*{line:for_loop} (\texttt{for (i=0;i<n;++i)}) and~\ref*{line:for_assignment} (\texttt{a[i] = nondet();}) is a maximal $analyzed$ trace.\closeExample

\end{example}

%

We will now revisit these concepts from an entirely different technique: lazy predicate abstraction~\cite{henzinger2002lazy}.

Lazy predicate abstraction, the algorithm used by BLAST ~\cite{henzinger2002lazy}, consists of two alternated phases.
The first phase generates, on-the-fly, a reachability tree whose nodes correspond to vertices of the Control Flow Automaton\footnote{A Control Flow Automaton, similar to a Control Flow Graph, captures the control flow of the program, where nodes correspond to locations and edges are labeled with statements.} (CFA) of the program.
This process goes on until exhaustively exploring the tree or until reaching a node that corresponds to an assertion failure.
Each node is associated with a predicate, initially $true$, that must hold for any path reaching that node and helps prune the successors that are not reachable.
If, and when, a node that represents an error is reached, the second phase deals with analyzing the potential counterexample to determine whether the path reaching the error node is feasible.
If the latter phase determines the potential counterexample is infeasible, the reachability tree is refined by strengthening the predicates associated to the appropriate nodes so that the path reaching the error node is pruned.
Lastly, when a counterexample is produced, it can be checked with a more precise analysis to ensure its feasibility.
However, if this additional check fails, the search is abandoned.

\begin{example}[Analyzed behaviors in lazy predicate abstraction]
    \label{ex:analyzedLazy}

    The code presented in Figure~\ref{fig:non_linear}, reproduced from the paper presenting Conditional Model Checking~\cite{beyer2011conditional,beyer2012conditional}, contains a non-linear safety property.
    As explained, the construction of the reachability tree will reach the error node and the second phase would attempt to verify the feasibility of the path leading to it.
    The feasibility check is usually implemented by creating a verification condition to be checked by an underlying SMT solver and, therefore, inherits the latter's limitations.
    In particular, SMT solvers usually cannot handle non-linear arithmetic and, instead, model multiplication as uninterpreted functions.
    Concretely, the SMT solver would not be able to prove the path infeasible.
    However, the subsequent counterexample feasibility check would prove the path is actually infeasible, causing the exploration to stop.
    
%
%
    Given the failure to analyze the assertion, the following is a maximal $analyzed$ trace: lines~\ref*{line:prefix1} (\texttt{int p = nondet();}),~\ref*{line:prefix2} (\texttt{if(p)}),~\ref*{line:frontier1} (\texttt{int x = 5;}),~\ref*{line:frontier2} (\texttt{int y = 6;}) and~\ref*{line:frontier3} (\texttt{int r = x * y;}).

    On the other hand, the \emph{then} branch of the \texttt{if}-statement can be successfully verified with this technique, since tracking the predicate \texttt{i < 1000000} suffices to prove the assertion always holds when execution leaves the \texttt{for}-loop.

    The successful verification of the \emph{then} branch would make the trace composed of lines~\ref*{line:prefix1} (\texttt{int p = nondet();}),~\ref*{line:prefix2} (\texttt{if (p)}) and~\ref*{line:safe} (\texttt{int i;}) a \emph{safe cone}. \closeExample

\end{example}

\begin{figure}
\begin{lstlisting}
int nondet();
int main() {
  %*\label{line:prefix1}*)int p = nondet();
  %*\label{line:prefix2}*)if (p) {
      %*\label{line:safe}*)int i;
      for (i = 0; i < 1000000; i++);
      assert(i >= 1000000);
  } else {
      %*\label{line:frontier1}*)int x = 5;
      %*\label{line:frontier2}*)int y = 6;
      %*\label{line:frontier3}*)int r = x * y;
      %*\label{line:frontier4}*)assert(r >= x);
  }
  return 0;
}
\end{lstlisting}
\caption{Non-linear arithmetic. Reproduced from CMC papers~\cite{beyer2011conditional,beyer2012conditional}.}
\label{fig:non_linear}
\end{figure}

\subsection{Execution Reports (ERs)}
\label{subsec:execution_reports}

We now formally define execution reports as an under-approximation of the set of \emph{safe cones} and $frontier$ traces. These definitions rely on two predicates ($analyzed?$ and $isSafeCone?$) whose definition depends greatly on the underlying verification technique used in the incomplete verification attempt. Consequently, in this section we simply provide properties that predicates $analyzed?$ and $isSafeCone?$ ought to satisfy. In the next section we ground the definition of these predicates for Abstract Reachability Tree based verification techniques~\cite{henzinger2002lazy}.

Examples~\ref{ex:analyzedBMC} and~\ref{ex:analyzedLazy} illustrate how the notions of $analyzed$ and \emph{safe cone} apply to diverse techniques.
We will capture these notions as predicates over traces in the following definitions.

Let the alphabet $\Sigma$ contain all statements in a program, making $\pi \in \Sigma^*$ a sequence of statements.

\begin{myprop}
    \label{prop_analyzed}
    The predicate $analyzed? : \Sigma^* \rightarrow Bool$ satisfies the following property, where $\cdot$ stands for concatenation:

        \label{prop:not_analyzed_implies_extension_not_analyzed} $\forall \pi, \pi' \in \Sigma^*. \; \neg analyzed?(\pi)  \rightarrow \neg analyzed?(\pi \cdot \pi') $
\end{myprop}

Property~\ref{prop:not_analyzed_implies_extension_not_analyzed} aims to formalize the notion of incrementality that we implicitly used throughout the examples.
Note that this property is logically equivalent to its contrapositive, that is, $analyzed?(\pi \cdot \pi') \rightarrow analyzed?(\pi)$, as expected of an incremental exploration.

\begin{myprop}
    The predicate $isSafeCone? : \Sigma^* \rightarrow Bool$ satisfies the following property, where $\cdot$ stands for concatenation:

        \label{prop:safe_cone} $\forall \pi, \pi' \in \Sigma^*. \; isSafeCone?(\pi) \rightarrow isSafeCone?(\pi \cdot \pi')$
\end{myprop}

Property~\ref{prop:safe_cone} reflects our notion of \emph{safe cone} as a trace reaching a fully analyzed portion of the behavior space.
Any trace extension will necessarily also be a \emph{safe cone}.

\begin{myprop}
    The predicate $isSafeCone? : \Sigma^* \rightarrow Bool$ satisfies the following property:

        \label{prop:safe_cone_implies_analyzed} $\forall \pi \in \Sigma^*. \; isSafeCone?(\pi) \rightarrow analyzed?(\pi)$
\end{myprop}

Property~\ref{prop:safe_cone_implies_analyzed} formalizes the connection between the two predicates, in particular how $isSafeCone?(\pi)$ subsumes \linebreak $analyzed?(\pi)$.

\begin{mydef}
    Given a trace $\pi \in \Sigma^*$ and a program $\mathcal{P}$, we introduce the following predicate:

    $feasible_{\mathcal{P}}(\pi)$ holds \emph{iff} there exists a concrete execution of the program $\mathcal{P}$ that executes $\pi$.
\end{mydef}

Given that we will always refer to a single program at a time, the system\hyp under\hyp analysis, we will omit the subscript.

\begin{myprop}
    \label{prop_analyzed_phi}
   Given $\varphi : \Sigma^* \rightarrow Bool$, a boolean predicate that captures the safety property of interest, the predicate $analyzed? : \Sigma^* \rightarrow Bool$ satisfies the following property:
    
        \label{prop:analyzed_safe} $\forall \pi \in \Sigma^*. \; analyzed?(\pi) \land \; feasible(\pi) \rightarrow \varphi(\pi) $
\end{myprop}

Property~\ref{prop_analyzed_phi} is at the core of interpreting $analyzed$ traces as providing safety assurances.
This property also holds for $isSafeCone?$ due to Property~\ref{prop:safe_cone_implies_analyzed}.
The predicate $feasible(\pi)$ in the antecedent places the focus of safety assurances on feasible traces, that is, traces that correspond to actual behaviors of the system-under-analysis. 

Recall that we provide properties that constrain the predicates $analyzed?$ and $isSafeCone?$ but not concrete definitions of these predicates as specific definitions depend on the underlying verification technique.
We now define the set of \emph{safe cones} and \emph{frontier traces} of an incomplete verification attempt.

\begin{mydef}
    \label{def_feasible_safe_cones}
The set $SafeCone$ is defined as follows:

    $SafeCone = \{\pi \cdot s | \pi \in \Sigma^*, s \in \Sigma, \neg isSafeCone?(\pi) \land feasible(\pi \cdot s) \land isSafeCone?(\pi \cdot s)\}$
\end{mydef}

\begin{mydef}
    \label{def_feasible_frontier_traces}
The set $Frontier$ is defined as follows:

$Frontier = \{\pi \cdot s | \pi \in \Sigma^*, s \in \Sigma, \ analyzed?(\pi) \land feasible(\pi \cdot s) \land \neg analyzed?(\pi \cdot s) \}$
\end{mydef}

Definitions~\ref{def_feasible_safe_cones} and~\ref{def_feasible_frontier_traces} are related to Properties~\ref{prop:safe_cone} and~\ref{prop_analyzed} respectively, since the incrementality of the analysis is key to the search for maximal $analyzed$ traces, as in the set $Frontier$, and minimal \emph{safe cone} traces, as in $SafeCone$. 

The set $SafeCone$ can provide safety assurances about the system, as in Example~\ref{ex:analyzedLazy}, where the \emph{then} branch of the \texttt{if}-statement has been fully verified.

Conversely, $Frontier$ can suggest shortcomings in the incomplete verification attempt.
For instance, in Example~\ref{ex:analyzedBMC}, the existence of a trace $\pi \in Frontier$ that did not even go past the initialization loop suggests an important part of the test harness was not sufficiently analyzed.

The conclusions obtained from inspecting both sets can be useful to assess the progress achieved throughout the incomplete verification attempt.

We anticipated the intuitive notion captured by these definitions in Examples~\ref{ex:analyzedBMC} and~\ref{ex:analyzedLazy}, but there is one important consideration that we omitted so far and now included in the definitions: feasibility.

Feasibility is relevant because, by definition, infeasible traces do not correspond to system behaviors.
Without feasibility guarantees, interpreting each trace would require careful analysis, because it could mislead a user into either increasing or decreasing her confidence in the system-under-analysis. 

\begin{mydef}
    \label{def:execution_report}
    An \emph{execution report} is a tuple $(S, F)$ where $S \subseteq SafeCone$ and $F \subseteq Frontier$.
\end{mydef}

Definition~\ref{def:execution_report} defines \emph{execution reports} as under\hyp approximations of the sets $SafeCone$ and $Frontier$, allowing empty sets as valid \emph{execution reports}.

The sets $SafeCone$ and $Frontier$ can grow quickly, making it extremely inefficient to compute the full sets.
Furthermore, some of the traces can be redundant, in a sense, if they only differ in a few statements, making it sensible to under-approximate.

Ideally, it would be desirable to characterize these under-approximation. However, we opted in this work for a notion of \emph{completeness} with respect to statements in the code that does not fully characterize the sets $S$ and $F$ but does not allow, in the general case, empty sets as valid \emph{execution reports}:
For both sets, $SafeCone$ and $Frontier$, we require that if a trace $\pi \in SafeCone$  (resp. $Frontier$) ending in a specific location $l$ exists, then there exists $\pi' \in S$ (resp. $F)$ and $\pi'$ also ends in $l$.
This completeness guarantee does not force extremely large sets of paths to be reported and loosely resembles a notion of coverage.
Our algorithm to generate \emph{execution reports} guarantees this completeness criterion.

\section{Reports for ART-based implementations}

Throughout this section we will explain how we generate Execution Reports for Abstract Reachability Tree (ART)-based ~\cite{henzinger2002lazy} techniques.

ARTs constitute a relevant intermediate data structure used in verification.
The variety of techniques implemented using ARTs makes them ideal for our proof-of-concept implementation.
ART-based implementations comprise a wide range of dissimilar techniques, encompassing lazy abstraction~\cite{henzinger2002lazy}, BMC~\cite{biere1999symbolic}, explicit value analysis~\cite{beyer2013explicit} and CEGAR variants of some of the previous~\cite{beyer2013explicit}, among other.

In order to explain how we generate Execution Reports for ART-based techniques, we first define $analyzed?$ and $isSafeCone?$ for these techniques in subsection~\ref{sec:preds-ART}.

We will use Assumption Automata, an existing machine\hyp readable abstract representation of ARTs, for our implementation.
In subsection~\ref{sec:AA} we briefly explain Assumption Automata and their two states most relevant to us, \texttt{TRUE} and \texttt{FALSE}, which we will use to obtain \emph{safe cones} and $frontier$ traces, respectively.

\begin{figure}
    \includegraphics[width=\columnwidth]{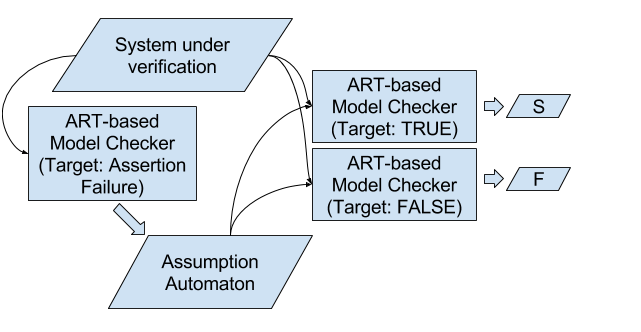}
\caption{Architecture of ER generation.}
\label{fig:architecture}
\end{figure}

Finally, in subsection~\ref{sec:algorithm} we will explain how we compute Execution Reports using an Assumption Automaton, produced by an earlier incomplete verification attempt, and the system\hyp under\hyp verification as input, as depicted in Figure~\ref{fig:architecture}.

\subsection{ARTs as intermediate data structures}
\label{sec:preds-ART}

%
%
We have discussed how the concepts of $analyzed$ and \emph{safe cone}, captured by predicates $analyzed?$ and $isSafeCone?$ respectively, apply to example techniques.
In this sub\hyp section we will instantiate these concepts to Abstract Reachability Trees (ARTs)~\cite{henzinger2002lazy}.

An ART is a tree whose nodes correspond to vertices of the CFA of a program and each node is associated to an element of an abstract domain.
In the case of BLAST, that abstract domain is a lattice of predicates.

ART-based algorithms consist of two phases, the first one comprises an incremental generation of the ART and the second phase involves a more thorough counterexample check.

For the purpose of stating which traces can be considered $analyzed$ we can ignore the abstract domain elements associated to each node.
We can then think of an ART as a tuple $(G, W, q_0, $\emph{covered}$)$ where $G = (S, \Sigma, \delta)$ is a graph, $W \subseteq S$ represents a wait list, $q_0 \in S$ is the initial state, $covered : S \rightarrow S$ captures subsumption between nodes, and $\delta : S \times \Sigma \rightarrow S$ is a transition function.
The graph $G$ captures the structure of the partially built ART.
$W$ is the wait list that contains elements to be analyzed in order to continue the construction of the ART.
The function $covered$ is necessary to represent subsumption between nodes but is not total.
We say that a node $e$ is covered by $e'$ \emph{iff} $covered(e) = e'$.
Analogously, we consider that a node $e$ is not covered \emph{iff} $covered(e)$ is undefined.
Successors of a subsumed node $e$ are not explored because the state space captured by $e$ is also represented by $covered(e)$.
Hence, subsumption is relevant to defining which parts of the state space can be considered explored.

To make the definitions easier to read, we will assume the following invariant holds for ARTs relative to $covered$:

\begin{myprop}

    \label{prop_covered}
    $\forall e \in S.\text{ such that }covered(e)\text{ is defined, then }\linebreak covered(covered(e))$ is not defined.
\end{myprop}

Informally, this means that no node is covered by another covered node.

In ART-based algorithms, a node $e$ covered by $e'$ need not be further analyzed because any error state found from $e$ will also be found from $e'$.
However, a sequence of statements that reaches $e$ might not be a safe trace if it is possible to reach an assertion failure from $e'$.
In other words, the sub-tree rooted in $e$ cannot be considered exhaustively built unless that is also the case for the sub-tree rooted in $e'$.
This observation is crucial to define what can be regarded as $analyzed$ or \emph{safe cone}.

We will extend $\delta$ as $\delta' : (S \cup \{ None \}) \times \Sigma \rightarrow S \cup \{ None \}$, with $None \notin S$ to make it total and resolve the $covered$ function transparently as follows:

\begin{align*}
    \delta'(q, s) =
    \begin{cases*}
	    \delta(q, s) & if $\delta(q, s)$ is defined and\\
	    	         & $covered(\delta(q,s))$ is not\\
        covered(\delta(q, s)) & if both $\delta(q, s)$ and\\
        		      & $covered(\delta(q, s))$ are defined\\
        None         & otherwise
    \end{cases*}
\end{align*}

Due to Property~\ref{prop_covered}, $\delta'(q, \pi)$ is never a covered node.

Moreover, we will adapt $\delta'$ to traces with $\hat{\delta'} : S \cup \{ None \} \times \Sigma^* \rightarrow S \cup \{ None \}$ as follows:
\begin{center}
    Given $q \in S \cup \{None\}$, $s \in \Sigma \text{ and } \pi \in \Sigma^*$:

	\[\begin{array}{c}
    \hat{\delta'}(q, s) = \delta'(q, s) \\
    \hat{\delta'}(q, s \cdot \pi) = \hat{\delta'}(\delta'(q, s), \pi)
	\end{array}
	\]
\end{center}

We will consider that $analyzed?(\pi)$ holds for a trace $\pi$ \emph{iff} no prefix of $\pi$ reaches a node in $W$ from the initial node.
That is:
\begin{equation*}
	analyzed?(\pi) \text{ \emph{iff} } \neg \exists \pi' \in \Sigma^*. isPrefix(\pi', \pi) \land \hat{\delta'}(q_0, \pi') \in W
\end{equation*}

Informally, we consider $\pi$ \emph{analyzed} when no prefix of $\pi$ reaches one of the states pending exploration, i.e. those in $W$.
The predicate $analyzed?$ is clearly monotonic, satisfying Property~\ref{prop_analyzed}: $\neg analyzed?(\pi)$ means there exists a prefix that reaches $W$, therefore any extension $\pi \cdot \pi'$ will also share that prefix.

Similarly, we will consider that $isSafeCone?(\pi)$ holds for a trace $\pi$ \emph{iff} any prefix of $\pi$ leads, from the initial node, to a sub-tree already exhaustively built.
Intuitively, a sub-tree is exhaustively built when none of its nodes are in $W$, the set containing states pending exploration.
More precisely:
\begin{gather*}
	isSafeCone?(\pi) \text{ \emph{iff} } \\
	\exists \pi' \in \Sigma^*. isPrefix(\pi', \pi) \land \forall \pi'' \in \Sigma^*. \hat{\delta'}(q_0, \pi' \cdot \pi'') \notin W
\end{gather*}
Analogously, the predicate $isSafeCone?$ is monotonic, satisfying Property~\ref{prop:safe_cone}: $isSafeCone?(\pi)$ means there exists a prefix from which $W$ is unreachable and consequently any extension $\pi \cdot \pi'$ will also share that prefix.


\myworries{TODO(rcastano): Should I discuss lazy abstraction example in detail, mentioning the ART (including Figure~\ref{ART_nonlinear})?}



\myworries{TODO(VICTOR): How do I talk about completeness with respect to locations without having it sound artificial?}

\subsection{Assumption Automata}
\label{sec:AA}

We  now briefly explain Assumption Automata~\cite{beyer2011conditional} because our implementation takes an Assumption Automaton as part of its input instead of ARTs.

An Assumption Automaton is essentially an abstraction of an ART where the elements of the abstract domain associated to each node are removed.
Additionally, the structure is compressed by collapsing sub-trees which have been entirely verified into a single node \texttt{TRUE} and every node covered by another node is merged with the latter.
Finally, nodes in the wait list are only connected to a single node \texttt{FALSE}.
An Assumption Automaton encodes the progress achieved throughout an earlier incomplete verification attempt in a machine-readable format~\cite{beyer2015witness}, which is the fundamental reason why we use it.

The definition of predicate $analyzed?(\pi)$ can also be stated in terms of Assumption Automata. The predicate holds \emph{iff} no prefix of $\pi$ reaches \texttt{FALSE}.
Once again, understanding an Assumption Automaton as a graph, the predicate is defined as follows:
\begin{gather*}
	analyzed?(\pi) \text{ \emph{iff} } \\
	\neg \exists \pi' \in \Sigma^*\text{ such that }isPrefix(\pi', \pi) \land \hat{\delta'}(q_0, \pi') = \texttt{FALSE}
\end{gather*}

Similarly, $isSafeCone?(\pi)$ holds iff a prefix of $\pi$ reaches the node \texttt{TRUE}.
That is:
\begin{gather*}
	isSafeCone?(\pi) \text{ \emph{iff} } \\
	\exists \pi' \in \Sigma^*\text{ such that }isPrefix(\pi', \pi) \land \hat{\delta'}(q_0, \pi') = \texttt{TRUE}
\end{gather*}

It is worth noting that, even though one Assumption Automaton can correspond to several ARTs, applying the predicates $isSafeCone?$ and $analyzed?$ to the former or to any of the latter will yield the same result.

\subsection{Generating reports for CPAchecker}
\label{sec:algorithm}

We built a proof-of-concept implementation, consisting of two verification tasks, capable of generating execution reports for ART-based techniques on top of CPAchecker.

\myworries{TODO(rcastano): Need more explaining of Assumption Automata?}



The input for our implementation is an Assumption Automaton and the original system-under-analysis.
Our output are the sets $S$ and $F$, composed of the counterexamples, i.e.\ traces, generated by two independent verification tasks, as shown in Figure~\ref{fig:architecture}.

We resort to the conceptual framework of Configurable Software Verification~\cite{beyer2007configurable} to formalize how our algorithm is parametric, allowing different reachability analyses to be used.
It is worth mentioning that the algorithm used to generate an Execution Report is in no way related to or restricted by the technique used for the original verification attempt, as long as the latter generates an Assumption Automaton.

Informally, we augment an existing ART-based algorithm by adding an Assumption Automaton state to the abstract domain element associated with each node.
That is, in Figure~\ref{fig:architecture}, the process that produces $F$ will attempt to find feasible traces that reach the state \texttt{FALSE} in the Assumption Automaton, whereas \texttt{TRUE} will be the target state in the case of the process generating $S$.

\myworries{TODO(rcastano): maybe illustrate with the Assumption Automaton example?}

We already explained the basics of ART-based algorithms, but some more detail is necessary to define our extension.
The framework of Configurable Software Verification allows us to define a Configurable Program Analysis (CPA) $\mathbb{P} = (D_{\mathbb{P}}, \rightsquigarrow_{\mathbb{P}}, \texttt{merge}_{\mathbb{P}}, \texttt{stop}_{\mathbb{P}})$ in terms of an abstract domain $D_{\mathbb{P}}$, a transfer relation $\rightsquigarrow_{\mathbb{P}}$, a merge operator $\texttt{merge}_{\mathbb{P}}$, and a termination check $\texttt{stop}_{\mathbb{P}}$.

Moreover, we can define a CPA as a composition of other CPA.
CPA composition formalizes how we can plug our set-specific analysis, e.g.\ $SafeCone$ or $Frontier$, into existing reachability analyses, such as explicit value analysis or predicate abstraction. 

A CPA, either simple or composite, can be analyzed with one of the several variants~\cite{beyer2007configurable,beyer2012algorithms} of the basic algorithm used in Configurable Software Verification, which we reproduced in Algorithm~\ref{alg}.

\begin{algorithm}[!ht]
\SetKwInOut{Input}{input}
\SetKwInOut{Output}{output}
\Input{
    A configurable program analysis $\mathbb{P} = (D, \rightsquigarrow, \texttt{merge}, \texttt{stop})$,
    a set \texttt{waitlist} of elements of $E$, denoting the set of elements of the semi-lattice of $D$,
    a set \texttt{reached} of reachable abstract states.
}
 \Output{An updated \texttt{reached} and \texttt{waitlist}.}
%

 \While{$\texttt{waitlist} \neq \emptyset$}{

     pop $e$ from \texttt{waitlist}
     
     \For{each $e'$ with $e \rightsquigarrow e'$}{

         \For{each $e'' \in \texttt{reached}$}{
             // Combine with existing abstract state

             $e_{new}$ := \texttt{merge}($e'$, $e''$)

             \If{$e_{new} \neq e''$}{

                 \texttt{waitlist} := (\texttt{waitlist} $\cup \{e_{new} \}) \setminus \{ e'' \}$

                 \texttt{reached} := (\texttt{reached} $\cup \{e_{new}\}) \setminus \{e''\}$
             }
         }

         \If{$\neg \texttt{stop}(e', \texttt{reached}$)}{

             \texttt{waitlist} := \texttt{waitlist} $\cup \; \{e'\}$

             \texttt{reached} := \texttt{reached} $\cup \; \{e'\}$

             \If{\texttt{isTargetState}($e'$)}{ \label{alg:added1}

                 \Return{(\texttt{reached}, \texttt{waitlist})} \label{alg:added2}
             }
         }
     }

     \Return{(\texttt{reached}, $\emptyset$)}
 }
 \caption{Basic algorithm (from Configurable Software Verification~\cite{beyer2007configurable,beyer2012algorithms})}
\label{alg}
\end{algorithm}
Algorithm~\ref{alg} differs only slightly from the one in the original presentation of Configurable Software Verification~\cite{beyer2007configurable}.
We added lines~\ref{alg:added1} and~\ref{alg:added2}, because we want to stop the exploration and return as soon as a target state is found.
We also consider the sets \texttt{waitlist} and \texttt{reached} as inputs, making the core analysis more amenable to extensions~\cite{beyer2013explicit,beyer2012algorithms}, such as CEGAR~\cite{clarke2003counterexample} or finding multiple counterexamples, which requires a similar approach. 

The check performed in line~\ref{alg:added1}, \texttt{isTargetState($e'$)}, as mentioned, will verify whether \texttt{TRUE} (respectively \texttt{FALSE}) is part of the composite state $e'$.
\texttt{TRUE} captures the portions of the state space which have been exhaustively verified, whereas \texttt{FALSE} corresponds to nodes in the ART pending analysis at the time the verification attempt was interrupted.
\myworries{TODO(rcastano): make sure the naming is consistent, I'm referring to safe traces here.}
In other words, the check evaluates whether any path $\pi$ leading to $e'$  satisfies $isSafeCone?(\pi)$ (resp. $\neg analyzed?(\pi)$).
Any prefix of $\pi$ will necessarily satisfy the negation of the check, otherwise it would have triggered the generation of a counterexample.
Therefore, if the check \texttt{isTargetState($e'$)} is positive, as long as $feasible(\pi)$ holds, $\pi \in SafeCone$ (resp. $\pi \in Frontier)$ since Property~\ref{def_feasible_safe_cones} (resp. Property~\ref{def_feasible_frontier_traces}) is satisfied by any such trace $\pi$. 

If such a trace $\pi$ is found, phase two collects $\pi$ if its feasibility is confirmed.
Regardless, Algorithm~\ref{alg} is executed from where it left off, since the full internal representation, the sets \texttt{reached} and \texttt{waitlist}, was returned at the end of the previous call.
This alternation between phase one and two produces a number of traces which will constitute $S$ (resp. $F$) in the \emph{execution report}.

\myworries{TODO(rcastano): How to discuss termination without getting into the details about error states, edges of \texttt{TRUE} and \texttt{FALSE}?? }

In order to preserve the properties of ER, we require the underlying analysis to be precise, that is, it does not produce spurious counterexamples.
Precise variants of different analyses, such as predicate abstraction and explicit value analysis, have been expressed as CPA~\cite{beyer2012algorithms,beyer2013explicit}.
The actual algorithms for these techniques are based on Algorithm~\ref{alg} and contain additional modifications but in both cases the composition with other CPA is still supported.

Let's now define our CPA $\mathbb{P}(A) = (D_{\mathbb{P}}, \rightsquigarrow_{\mathbb{P}}, \texttt{merge}_{\mathbb{P}}, \texttt{stop}_{\mathbb{P}})$
with $D_{\mathbb{P}}$ based on the flat lattice for the set of all the states in the Assumption Automaton $A$ taken as input.
For the transfer relation, $e \xrsquigarrow{stm}_{\mathbb{P}} e'$ if there exists a transition labeled $stm$ from state $e$ to state $e'$ of the Assumption Automaton $A$ taken as input and $e \xrsquigarrow{stm}_{\mathbb{P}} \bot$ otherwise.
Finally, $\texttt{merge}_{\mathbb{P}} = \texttt{merge}^{sep}$ and $\texttt{stop}_{\mathbb{P}} = \texttt{stop}^{sep}$, where $\texttt{merge}^{sep}(e, e') = e'$ and $\texttt{stop}^{sep}(e, R) = (\exists e' \in R: e \sqsubseteq e')$.
The operator $\texttt{merge}_{\mathbb{P}}$ and $\texttt{stop}_{\mathbb{P}}$ affect the precision and performance of the analysis~\cite{beyer2007configurable}.
The proposed definitions for these operators aim to increase the former possibly at the cost of the latter.
In any case, the analysis can be made entirely precise by checking the feasibility before reporting any counterexample.
Therefore, we define these operators for completeness but other options are entirely possible and might be desirable for performance.

We are now ready to define a composite program analysis $\mathcal{C} = (\mathbb{A}, \mathbb{P}, \rightsquigarrow_{\times}, \texttt{merge}_{\times}, \texttt{stop}_{\times})$, where $\mathbb{P}$ is the one just defined and $\mathbb{A}$ is an arbitrary analysis.
We will use $\texttt{merge}_{\times} = \texttt{merge}^{sep}$, $\texttt{stop}_{\times} = \texttt{stop}^{sep}$ and define the transfer relation $\rightsquigarrow_{\times}$ such that $(l, r) \xrsquigarrow{g}_{\times} (l', r')$ iff $l \xrsquigarrow{g}_{\mathbb{A}} l'$ and $r \xrsquigarrow{g}_{\mathbb{P}} r'$.

In order to satisfy the completeness guarantees with respect to locations in the program that we put forth in sub-section~\ref{subsec:execution_reports}, we only require that the underlying analysis does not merge different location states.
Neither of the analyses we tried, lazy predicate analysis nor explicit value analysis, do.

It is worth emphasizing that the predicates $analyzed?$ and $isSafeCone?$ only play a conceptual role to make sure our implementation preserves the informal semantics of $frontier$ traces and \emph{safe cones} discussed in Section ~\ref{sec:motivation}.
We merely mapped the intuitions to the specific case of ART-based algorithms and computed the sets $S$ and $F$ directly from the Assumption Automaton.

\section{Evaluation}

This section aims to evaluate if our approach is capable of generating informative ERs within a reasonable time (with respect to the time budget invested in the original verification attempt).
We analyze the performance of our proof-of-concept implementation with a set of standard benchmark instances.
We also discuss the insights that we extracted from the output on this set of benchmarks.

All the files necessary to reproduce the experiments are available online: \url{https://github.com/rcastano/cpachecker-1/tree/submission_ase}.

\subsection{Performance}


\begin{table*}
    \centering
	\begin{tabular}{l|c|c|c|c|c|c|l}
 & \multicolumn{3}{c|}{Safe:  explicit from explicit-AA} & \multicolumn{3}{c|}{Safe:  predicate from explicit-AA} \\ 
 & \# traces & Full? & CPU time & \# traces & Full? & CPU time & \\
gigaset.BUG.c 	& - & \xmark 	& 927.44s 	& 1 & \checkmark 	& 417.08s 	&  \\
farsync.BUG.c 	& - & \xmark 	& 926.65s 	& 1 & \checkmark 	& 280.36s 	&  \\
loop.BUG.c 	& - & \xmark 	& 901.08s 	& - & \xmark 	& 135.17s 	&  \\
synclink\_gt.BUG.c 	& 1 & \checkmark 	& 883.54s 	& 2 & \checkmark 	& 534.85s 	&  \\
ppp\_generic.BUG.c 	& 1 & \checkmark 	& 894.03s 	& 1 & \checkmark 	& 855.15s 	&  \\
lirc\_imon.BUG.c 	& - & \xmark 	& 901.05s 	& - & \xmark 	& 911.77s 	&  \\
token\_ring.14.BUG.c 	& 19 & \checkmark 	& 25.93s 	& - & \xmark 	& 902.45s 	&  \\
transmitter.16.BUG.c 	& 7 & \checkmark 	& 16.43s 	& - & \xmark 	& 901.97s 	&  \\
toy.c 	& - & \xmark 	& 970.09s 	& 0 & \checkmark 	& 893.52s 	&  \\
 & &4/9 Full & & &5/9 Full & &  \\
\end{tabular}

	\caption{Complete results for S from an Assumption Automaton generated using Explicit Value analysis}
	\label{table:s_explicit}
\end{table*}

\begin{table*}
    \centering
	\begin{tabular}{l|c|c|c|c|c|c|l}
 & \multicolumn{3}{c|}{Safe:  explicit from predicate-AA} & \multicolumn{3}{c|}{Safe:  predicate from predicate-AA} \\ 
 & \# traces & Full? & CPU time & \# traces & Full? & CPU time & \\
toy1\_BUG.c 	& 0 & \checkmark 	& 14.15s 	& 0 & \checkmark 	& 52.47s 	&  \\
token\_ring.06.c 	& 0 & \checkmark 	& 48.40s 	& 0 & \checkmark 	& 272.59s 	&  \\
pipeline.c 	& 0 & \checkmark 	& 7.66s 	& 0 & \checkmark 	& 10.93s 	&  \\
token\_ring.09.BUG.c 	& - & \xmark 	& 901.05s 	& - & \xmark 	& 901.07s 	&  \\
token\_ring.04.c 	& 0 & \checkmark 	& 20.60s 	& 0 & \checkmark 	& 78.64s 	&  \\
token\_ring.05.c 	& 0 & \checkmark 	& 50.49s 	& 0 & \checkmark 	& 299.75s 	&  \\
token\_ring.08.c 	& 0 & \checkmark 	& 367.83s 	& - & \xmark 	& 901.09s 	&  \\
token\_ring.14.BUG.c 	& - & \xmark 	& 963.61s 	& - & \xmark 	& 901.07s 	&  \\
mem\_slave\_tlm.3.c 	& 0 & \checkmark 	& 8.72s 	& 0 & \checkmark 	& 27.59s 	&  \\
token\_ring.07.c 	& 0 & \checkmark 	& 108.09s 	& - & \xmark 	& 901.08s 	&  \\
mem\_slave\_tlm.5.c 	& 0 & \checkmark 	& 15.80s 	& 0 & \checkmark 	& 145.01s 	&  \\
mem\_slave\_tlm.4.c 	& 0 & \checkmark 	& 9.00s 	& 0 & \checkmark 	& 27.56s 	&  \\
kundu.c 	& 0 & \checkmark 	& 16.90s 	& 0 & \checkmark 	& 135.16s 	&  \\
pktcdvd.BUG.c 	& - & \xmark 	& 900.97s 	& 5 & \checkmark 	& 166.12s 	&  \\
toy.c 	& 0 & \checkmark 	& 13.91s 	& 0 & \checkmark 	& 97.82s 	&  \\
token\_ring.03.c 	& 1 & \checkmark 	& 16.43s 	& 0 & \checkmark 	& 68.03s 	&  \\
 & &13/16 Full & & &12/16 Full & & \\
\end{tabular}

	\caption{Complete results for S from an Assumption Automaton generated using Predicate analysis}
	\label{table:s_predicate}
\end{table*}

\begin{table*}
    \centering
	\begin{tabular}{l|c|c|c|c|c|c|l}
 & \multicolumn{3}{c|}{Frontier:  explicit from explicit-AA} & \multicolumn{3}{c|}{Frontier:  predicate from explicit-AA} \\ 
 & \# traces & Full? & CPU time & \# traces & Full? & CPU time & \\
gigaset.BUG.c 	& - & \xmark 	& 930.02s 	& - & \xmark 	& 917.01s 	&  \\
farsync.BUG.c 	& - & \xmark 	& 918.63s 	& 13 & \checkmark 	& 98.98s 	&  \\
loop.BUG.c 	& - & \xmark 	& 937.49s 	& 4 & \checkmark 	& 153.23s 	&  \\
synclink\_gt.BUG.c 	& - & \xmark 	& 917.86s 	& - & \xmark 	& 1000.97s 	&  \\
ppp\_generic.BUG.c 	& - & \xmark 	& 901.01s 	& 30 & \checkmark 	& 94.69s 	&  \\
lirc\_imon.BUG.c 	& - & \xmark 	& 901.05s 	& - & \xmark 	& 912.70s 	&  \\
token\_ring.14.BUG.c 	& 4 & \checkmark 	& 12.65s 	& 4 & \checkmark 	& 16.58s 	&  \\
transmitter.16.BUG.c 	& 10 & \checkmark 	& 19.59s 	& 10 & \checkmark 	& 22.80s 	&  \\
toy.c 	& - & \xmark 	& 901.02s 	& - & \xmark 	& 931.81s 	&  \\
 & &2/9 Full & & &5/9 Full & & \\
\end{tabular}

	\caption{Complete results for F from an Assumption Automaton generated using Explicit Value analysis}
	\label{table:f_explicit}
\end{table*}

\begin{table*}
    \centering
	\begin{tabular}{l|c|c|c|c|c|c|l}
 & \multicolumn{3}{c|}{Frontier:  explicit from predicate-AA} & \multicolumn{3}{c|}{Frontier:  predicate from predicate-AA} \\ 
 & \# traces & Full? & CPU time & \# traces & Full? & CPU time & \\
toy1\_BUG.c 	& 10 & \checkmark 	& 20.44s 	& 23 & \checkmark 	& 133.11s 	&  \\
token\_ring.06.c 	& 153 & \checkmark 	& 83.51s 	& - & \xmark 	& 910.96s 	&  \\
pipeline.c 	& 1 & \checkmark 	& 8.93s 	& 1 & \checkmark 	& 19.71s 	&  \\
token\_ring.09.BUG.c 	& - & \xmark 	& 1000.85s 	& - & \xmark 	& 901.10s 	&  \\
token\_ring.04.c 	& 29 & \checkmark 	& 37.28s 	& 136 & \checkmark 	& 527.54s 	&  \\
token\_ring.05.c 	& 85 & \checkmark 	& 78.29s 	& - & \xmark 	& 900.90s 	&  \\
token\_ring.08.c 	& 871 & \checkmark 	& 628.35s 	& - & \xmark 	& 905.49s 	&  \\
token\_ring.14.BUG.c 	& - & \xmark 	& 1000.78s 	& - & \xmark 	& 939.53s 	&  \\
mem\_slave\_tlm.3.c 	& 5 & \checkmark 	& 14.97s 	& 5 & \checkmark 	& 25.82s 	&  \\
token\_ring.07.c 	& 337 & \checkmark 	& 188.40s 	& - & \xmark 	& 910.68s 	&  \\
mem\_slave\_tlm.5.c 	& 14 & \checkmark 	& 45.17s 	& 18 & \checkmark 	& 302.88s 	&  \\
mem\_slave\_tlm.4.c 	& 6 & \checkmark 	& 16.07s 	& 6 & \checkmark 	& 27.05s 	&  \\
kundu.c 	& 15 & \checkmark 	& 32.30s 	& - & \xmark 	& 292.07s 	&  \\
pktcdvd.BUG.c 	& - & \xmark 	& 901.07s 	& 39 & \checkmark 	& 274.03s 	&  \\
toy.c 	& 10 & \checkmark 	& 19.93s 	& 23 & \checkmark 	& 137.26s 	&  \\
token\_ring.03.c 	& 2 & \checkmark 	& 17.29s 	& 3 & \checkmark 	& 61.84s 	&  \\
 & &13/16 Full & & &9/16 Full & & \\
\end{tabular}

	\caption{Complete results for F from an Assumption Automaton generated using Predicate analysis}
	\label{table:f_predicate}
\end{table*}


We evaluated our algorithms using the families SystemC and DeviceDrivers of the SV-COMP~\cite{DBLP:conf/tacas/Beyer16} set of benchmarks, which were used previously to evaluate CMC.
Our experiments consisted of two different phases: 1) a verification attempt with a predefined time limit of 900 seconds; 2) production of ERs for the instances which ended the first phase with inconclusive results.

We used an Ubuntu 16.04 system equipped with an Intel\textsuperscript{\textregistered} Core\texttrademark~i7-3770 CPU clocked at 3.40GHz with 16GB of DDR3 memory for the experiments in a system without a swap partition running.
We used BenchExec~\cite{beyer2015benchmarking} to run the experiments and only allowed access to a single CPU core and 12GB of RAM.


\begin{table*}
    \centering
 		\begin{tabular}{l|r|r|r|r}
 		Verification & explicit    & predicate    & explicit     & predicate \\
 		ER           & predicate   & predicate    & explicit     & explicit \\
 		\hline\hline
$F$                  & 55.56 \%    & 50 \%        & 22.22 \%     & 75 \% \\
$S$                  & 22.22 \%    & 75 \%        & 22.22 \%     & 81.25 \% \\
				\hline\hline
	\end{tabular}
     \caption{Percentage of instances producing a full ER within 50\% of the original verification time (\SI{900}{s})
 		}
        \label{table:50p}
\end{table*}

    \begin{figure}
	    \centering
    \includegraphics[keepaspectratio,max width=.66\columnwidth]{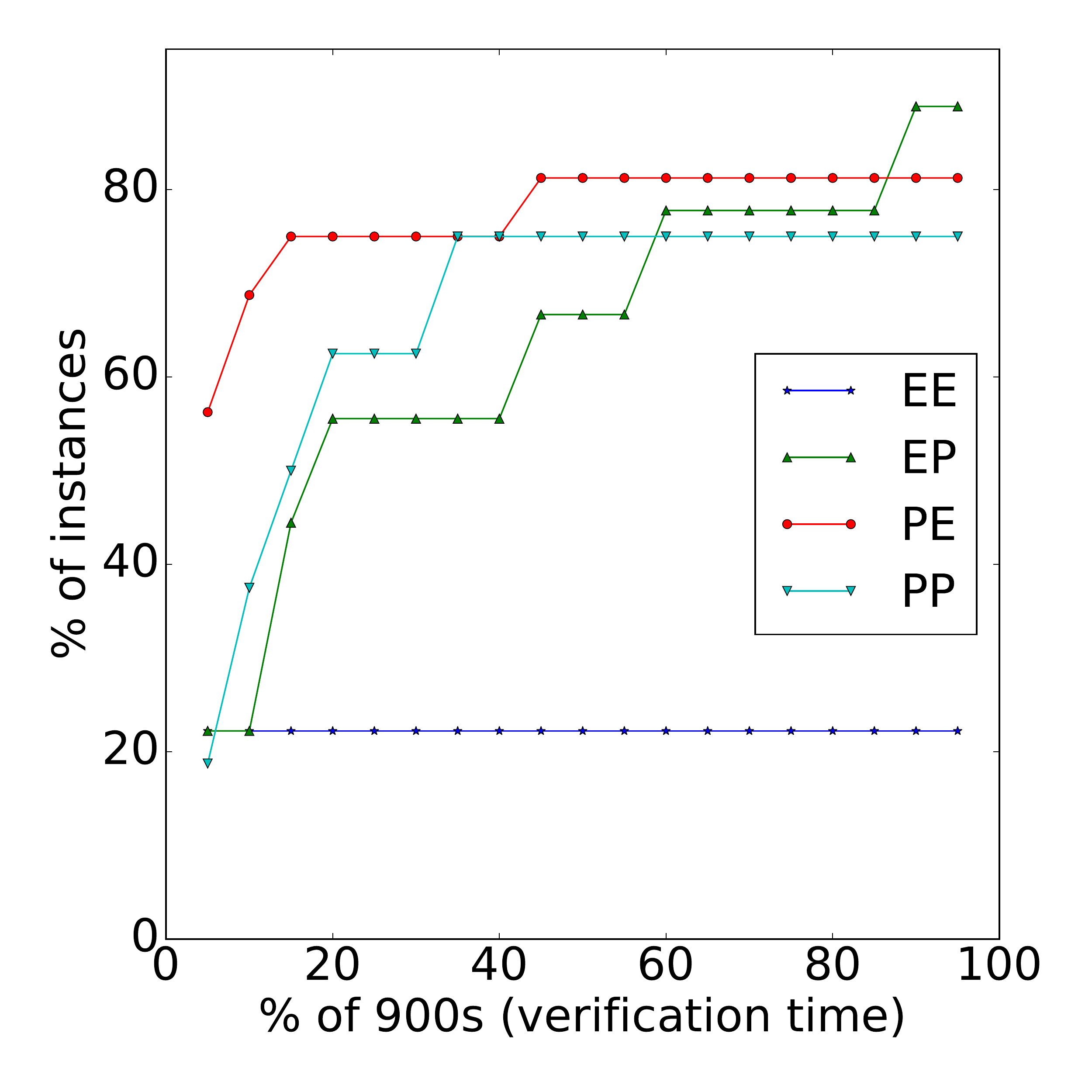}
    \caption{Time until first element of $S$ or $F$ is produced}
    \label{fig:one-or-more-other-than-explored}
\end{figure}

We ran the verification phase over all the SV-COMP instances included in the reproduction package\footnote{Available online: \url{https://www.sosy-lab.org/~dbeyer/cpa-cmc/}} of the original CMC presentation~\cite{beyer2012conditional}.
These benchmarks include instances from the sets SystemC and DeviceDrivers, comprising a total of 68 instances.
We used explicit value analysis (\texttt{EV}) and predicate abstraction (\texttt{PA}), leaving 9 and 16 instances, respectively, with inconclusive results.


In the second phase, for each of these executions interrupted due to reaching the time limit, we generate the corresponding ER.
To do so, we also experimented with both techniques, leading to 4 combinations of techniques, namely, \texttt{EV}-\texttt{EV}, \texttt{EV}-\texttt{PA}, \texttt{PA}-\texttt{EV} and \texttt{PA}-\texttt{PA}, where \texttt{EV}-\texttt{PA} denotes using \texttt{EV} for the first phase and \texttt{PA} for ER generation.
For each combination, we considered two configurations, one that finds a single trace for each of $S$ and $F$, to assess how early the second phase can produce output, and another one where we configure CPAchecker to continue listing counterexamples until exhausting its abstract state space, to determine the performance when generating the components to completion.\footnote{It is worth noting that a full ER component will satisfy the completeness guarantee with respect to locations in the program but will most likely not contain all valid traces.}

\Cref{table:s_explicit,table:s_predicate,table:f_explicit,table:f_predicate} include the full results of the second phase when running to completion.
The number of traces is ommitted for instances that did not yield conclusive results (for instance due to reaching the time limit or limitations of the analysis), as indicated with the dash.
The column labeled ``Full?'' indicates with a checkmark (\checkmark) when the component (either S or F) was generated to completion within the time limit.

We will mostly focus on the number of full components generated, that is, the number of instances for which the component, either S or F, is generated to completion within the allotted time, as indicated by the fraction below the second column for each combination of techniques.

\Cref{table:s_predicate,table:f_predicate} show that both generating the component S and the component F to completion seems possible for a significant number of instances when the Assumption Automaton was produced by lazy predicate abstraction.
Furthermore, using either lazy predicate abstraction again (right) or explicit value (left) for the generation of S shows similar performance results, with a slightly better performance of the latter.

In the case of F, there seems to be a greater difference in performance when using explicit value analysis (left) with respect to using lazy predicate abstraction (right) again for ER generation.

\Cref{table:f_predicate} suggests that using a different technique to generate the F component than the one used for verification could yield greater performance.
This pattern also shows up in \Cref{table:f_explicit}, where using lazy predicate abstraction (right) for the ER generation phase outperformed using explicit value analysis for an Assumption Automaton generated using explicit value analysis.

\Cref{table:s_explicit,table:f_explicit} show a lower number of instances for which the components were produced to completion within the allotted time, compared to the case when the Assumption Automaton was produced after an initial verification attempt using lazy predicate abstraction.
This could be explained by the size of Assumption Automata produced when using explicit value analysis, which is usually significantly larger than those produced with predicate analysis.

We consider these results encouraging in general, taking into account that the instances considered are those for which the initial verification attempt already exceeded the time limit.

Table~\ref{table:50p} aggregates these results and shows the percentage of instances for which a full ER was generated within 50\% (\SI{450}{s}) of the original verification time (\SI{900}{s}).
This table essentially measures how often the ideal condition holds, that is: both S and F are produced to completion within a fraction of the original verification time.
While the first, second and last column, in our opinion, suggest that standard techniques can achieve good performance and generate full ERs for widely dissimilar verification techniques, the third column shows that producing full ERs can, in some cases, demand significant computation, relative to the original verification time.
The latter finding comes as no surprise, since our approach to generating ERs constitutes a verification task in itself.
Nevertheless, it seems clear to us that consistently generating full ERs requires further work.

However, we are also interested in understanding how fast, relative to the original verification time, our implementation starts generating output.
Figure~\ref{fig:one-or-more-other-than-explored} shows the percentage of instances for which an element of either $F$ or $S$ is produced.
Figure~\ref{fig:one-or-more-other-than-explored} suggests that a fraction of the original verification time can be sufficient to start generating traces.

We consider these performance results encouraging for a proof-of-concept implementation and discuss some promising ideas for further improvement in Section~\ref{sec:conclusions}.

\subsection{Number of traces}

We mentioned a completeness criterion with respect to the statements of the system\hyp under\hyp verification, but this only provides a lower bound in the number of traces.
\Cref{table:f_explicit,table:f_predicate,table:s_explicit,table:s_predicate} show the number of traces that each component (S and F) contains.

The numbers in \Cref{table:s_predicate} prompted us to investigate the cause for the lack of \emph{safe cones}.
The underlying reason seems to be that reverse post\hyp order is used as the traversal strategy for lazy predicate abstraction.
This strategy heavily deprioritizes nodes with no successors in the CFA.
In particular, a final return statement is not analyzed unless no other option exists.
This exploration, unless additional specific configuration options are used, causes the analysis to seldom produce \emph{safe cones}.

This detailed inspection prompted by \Cref{table:s_predicate} allowed us to gain insights on the inner workings of the technique.
Moreover, wider availability of Execution Reports could provide an incentive for tool implementers to make implementation decisions that lead to richer intermediate results, which would be reflected in the corresponding Execution Reports.

\Cref{table:s_predicate,table:f_explicit} show that in many cases the number of traces produced remains manageable: within the few dozen and in many cases less than 5.
We consider these number encouraging, since little effort on behalf of the user would be required to inspect the traces and assess their usefulness.

However, \Cref{table:f_predicate} shows that for a few cases, the number of traces produced can exceed the hundreds.
In these cases, depending on the particular instance, it could be possible to group traces or inspect only a few, in both cases still gaining non-trivial information about the incomplete verification attempt with minimal effort.

\subsection{Discussion}


This section aims to shed light on the insights that can be extracted from ERs in practice.
With this goal, we manually inspected the Execution Reports produced for the families SystemC and DeviceDrivers of the SV-COMP benchmark and also for combined instances used to validate CMC~\cite{beyer2012conditional} and gained some anecdotal insights worth discussing.

Extracting insights from the ERs required knowledge about the instances.
However, a brief inspection of the code, guided by the output of our tool, yielded the necessary information.
The following features appeared repeatedly throughout the benchmark instances.

In all instances where the $Frontier$ component was finished we identified seemingly relevant behaviors that had not been analyzed.
More precisely, both SystemC and DeviceDrivers instances consist of a main while-loop with a non-deterministic termination condition and a set of methods from which one is chosen also non-deterministically and subsequently called.
In all of the instances the $Frontier$ component contained traces that did not even reach the end of the first loop iteration.
We exemplified this sort of output in Example~\ref{ex:analyzedBMC}, where the initialization loop could not be entirely analyzed under certain non-deterministic choices.
These insights contrast starkly with a coverage metric provided by CPAchecker.
Contrary to the indications of insufficient analysis reflected in the \emph{execution report}, the coverage metric provided by CPAchecker reported over 85\% line coverage for 11 out of the 16\myworries{(TODO(rcastano) check these numbers!)} instances when applying lazy abstraction.
This type of feedback could also be used to better understand the effects of either resource bounds or iteration bounds on the progress of the exploration.

	\begin{figure}
	\begin{lstlisting}
	void main () {
	int x = nondet();
	if (x) main0 ();
	if (!x) main1 ();
	}
	\end{lstlisting}
	\caption{Combined instance}
	\label{fig:combined}
	\vspace{-2.1em}
	\end{figure}

Generating insight from an element $e \in S$ requires not only knowledge of the instance but also understanding how $e$ constrains the execution from that point on.
For some instances, further research is needed to properly evaluate the relevance of each element $e \in S$.
However, for the combined instances, it was fairly easy to interpret ERs and they showed a relevant set of behaviors were safe.
These instances have the structure shown in Figure~\ref{fig:combined}, where method \texttt{main0} corresponds to the main method of a SystemC instance and \texttt{main1} corresponds to the main method of a DeviceDrivers instance.
Much like the setting we showed in Example~\ref{ex:analyzedLazy}, where one branch of an \texttt{if}-statement was easier to verify than the other, in this case \texttt{main0} is significantly less challenging, for a specific technique, than \texttt{main1}.
In these examples, $S$ would contain the trace \texttt{int x = nondet();}, \texttt{if(x)}, \texttt{main0()}.
The trace only constrains the value of \texttt{x}, which has to be positive, but \texttt{x} is a local variable, therefore \texttt{main0} is absolutely unconstrained and has been completely verified.

\section{Related Work}
Conditional model checking ~\cite{beyer2012conditional} (CMC) is an approach where model checkers are extended to produce results even when the verification run could not be completed successfully.
The output, in its general form, is a condition under which the program can be safely run.
CPAchecker~\cite{beyer2007configurable} instantiates CMC  by generating an Assumption Automaton.
We use this implementation to produce the ERs.
ERs introduce the notions of \emph{frontier} and \emph{safe cone} to characterize the state space denoted by structures like Assumption Automata. 
The lack of any feasibility guarantees in Assumption Automata has significant consequences.
A user might be misled by the size of the Assumption Automaton since a vast Automaton might correspond to a minuscule number of concrete feasible executions and also the other way around.
We overcome this limitation by formally characterizing the properties of our output and adding explicit feasibility guarantees.

A slightly different approach~\cite{christakis2012collaborative,DBLP:conf/icse/Christakis0W16} to providing feedback of partial verification results consists of a language extension to be used to annotate assumptions made during verification. 
This extension can be used to annotate the code and explicitly state conditions under which the program is guaranteed to run safely.
These annotations are especially well-suited for local assumptions, sometimes used during manual verification, and for uniform assumptions, such as the absence of integer overflows, which are not affected by the context in which they occur.
However, these annotations are not well-suited to state assumptions made by some techniques, such as those based in unrolling loops~\cite{christakis2012collaborative} or those tackled by Assumption Automata, and as such, they are incomparable to our approach, which can provide value in these cases.
One of the use cases of these annotations is to complement the verification efforts and produce a small test suite.
This idea of testing to complement earlier verification efforts was later replicated ~\cite{czech2015just} using Assumption Automata as input.

A recent extension of the Dafny IDE~\cite{leino2010dafny,DBLP:conf/tacas/ChristakisL0W16} provides, as one of its many features, hints about parts of a specification that might  cause timeouts.
However, given the modular nature of the tool and the sort of specifications shown as examples, the approach tackles a different problem than ours and the feature might not be applicable in our setting.

There is also previous work on quantifying partial model checker explorations based on usage profiles~\cite{pavese2010my}.
These estimations are based on abstract models of behavior and heavily depend on the provided usage profile.
A similar approach~\cite{filieri2013reliability} works applying symbolic execution over the source code of a system, without the need for an abstract model, but in this case, the implementation requires finite domains for all input variables, as well as a usage profile for each of them.
Reliability as they define it can be hard to interpret and, once again, could be extremely sensitive to the usage profile provided.
Both techniques can be used in conjunction with ours providing different value.

The modeling language Alloy~\cite{jackson2012software} enables its users to specify structural properties.
An extension of the Alloy Analyzer~\cite{d2010alloy+} highlights parts of the specification that are ``problematic'' or ``hard'', in the authors' own words, by monitoring the activity of variables and clauses in the underlying SAT solver.
The output is inherently heuristic, in contrast, our technique and proposed implementation provide strong guarantees backed by a formal definition of the semantics of the output generated.
The ideas behind the Alloy Analyzer extension could also be applied in conjunction with our techniques to provide additional information.

Our work is heavily influenced by the ideas and tool support of witness validation~\cite{beyer2015witness,beyer2015software}, which we leverage as a machine-readable representation of exploration progress.
However, we aim at enabling a richer manual interaction whereas that line of work also attempts to increase tool automation and reduce the need for manual inspection.

\section{Conclusions and future work}
\label{sec:conclusions}
Software model checking is already capable of handling industrial instances and produce valuable results.
However, some instances still remain intractable for full verification.
Our work provides users with a different way to observe the progress achieved during an incomplete verification attempt by producing an execution report (ER).

We formulated the concepts of $analyzed$ and \emph{safe cone} traces and formalized the notion of ERs.
We also discussed a proof-of-concept implementation to generate ERs and subsequently evaluated it both qualitatively and quantitatively with benchmark instances.

One line of research we plan to follow involves exploring ways to abstract the traces included in the execution reports for easier visualization and understanding. 
In this setting, it could be useful to define both existential and universal semantics for abstract traces in the sense that some property applies to the some or every concretization of these abstractions.
For instance, guaranteeing that every concretization has been analyzed, could be useful and would pose interesting challenges.
In a similar line, to better assess the relevance of an element $\pi \in S$, we need to devise meaningful views and metrics of the possible continuations of $e$, e.g.\ statement coverage metrics of the cone defined by $\pi$.

We would also like to analyze how our technique performs for a specific verification technique and varying time limits.

It would be desirable to conduct a user study to assess the effectiveness of our approach once more competing output representations become available.

We intend to look into alternative usages of our output.
For example, elements of an \emph{execution report} can be leveraged as additional input to choose the right algorithm~\cite{tulsian2014mux} to proceed after an incomplete verification attempt.

As mentioned in Section~\ref{sec:motivation}, the partition of system behaviors into the subsets proposed is also applicable to verification techniques beyond those already implemented using ARTs: Corral, DSE, techniques based on inlining or loop unrolling among other.
Several other verification systems already produce specification violation witnesses~\cite{beyer2015software} in a machine-readable format closely related to that of Assumption Automata.
Producing Assumption Automata for incomplete executions of many of these software model checkers seems possible, making our \emph{execution reports} immediately available to them.
Making the necessary changes to these tools and evaluating our approach could bring new insights and challenges worth looking into.

We also consider looking into performance improvements in the ER generation phase, for example using techniques that trade off soundness for efficiency.
More precisely, instead of using a verification tool for ER generation, we could use simulation or graph exploration techniques that are not necessarily exhaustive.

\bibliographystyle{abbrv}
\bibliography{related}

\begin{thebibliography}{10}

\bibitem{ball2011decade}
T.~Ball, V.~Levin, and S.~K. Rajamani.
\newblock A decade of software model checking with slam.
\newblock {\em Communications of the ACM}, 54(7):68--76, 2011.

\bibitem{beyer2015software}
D.~Beyer.
\newblock Software verification and verifiable witnesses.
\newblock In {\em Tools and Algorithms for the Construction and Analysis of
  Systems}, pages 401--416. Springer, 2015.

\bibitem{DBLP:conf/tacas/Beyer16}
D.~Beyer.
\newblock Reliable and reproducible competition results with benchexec and
  witnesses (report on {SV-COMP} 2016).
\newblock In Chechik and Raskin \cite{DBLP:conf/tacas/2016}, pages 887--904.

\bibitem{beyer2015witness}
D.~Beyer, M.~Dangl, D.~Dietsch, M.~Heizmann, and A.~Stahlbauer.
\newblock Witness validation and stepwise testification across software
  verifiers.
\newblock In {\em Proceedings of the 2015 10th Joint Meeting on Foundations of
  Software Engineering}, pages 721--733. ACM, 2015.

\bibitem{beyer2011conditional}
D.~Beyer, T.~A. Henzinger, M.~E. Keremoglu, and P.~Wendler.
\newblock Conditional model checking.
\newblock {\em arXiv preprint arXiv:1109.6926}, 2011.

\bibitem{beyer2012conditional}
D.~Beyer, T.~A. Henzinger, M.~E. Keremoglu, and P.~Wendler.
\newblock Conditional model checking: a technique to pass information between
  verifiers.
\newblock In {\em Proceedings of the ACM SIGSOFT 20th International Symposium
  on the Foundations of Software Engineering}, page~57. ACM, 2012.

\bibitem{beyer2007configurable}
D.~Beyer, T.~A. Henzinger, and G.~Th{\'e}oduloz.
\newblock Configurable software verification: Concretizing the convergence of
  model checking and program analysis.
\newblock In {\em Computer Aided Verification}, pages 504--518. Springer, 2007.

\bibitem{beyer2013explicit}
D.~Beyer and S.~L{\"o}we.
\newblock Explicit-state software model checking based on cegar and
  interpolation.
\newblock In {\em International Conference on Fundamental Approaches to
  Software Engineering}, pages 146--162. Springer, 2013.

\bibitem{beyer2015benchmarking}
D.~Beyer, S.~L{\"o}we, and P.~Wendler.
\newblock Benchmarking and resource measurement.
\newblock In {\em Model Checking Software}, pages 160--178. Springer, 2015.

\bibitem{beyer2012algorithms}
D.~Beyer and P.~Wendler.
\newblock Algorithms for software model checking: Predicate abstraction vs.
  impact.
\newblock {\em FMCAD 2012 Formal Methods in Computer--Aided Design}, page 106.

\bibitem{biere1999symbolic}
A.~Biere, A.~Cimatti, E.~M. Clarke, M.~Fujita, and Y.~Zhu.
\newblock Symbolic model checking using sat procedures instead of bdds.
\newblock In {\em Proceedings of the 36th annual ACM/IEEE Design Automation
  Conference}, pages 317--320. ACM, 1999.

\bibitem{cadar2013symbolic}
C.~Cadar and K.~Sen.
\newblock Symbolic execution for software testing: three decades later.
\newblock {\em Communications of the ACM}, 56(2):82--90, 2013.

\bibitem{DBLP:conf/tacas/2016}
M.~Chechik and J.~Raskin, editors.
\newblock {\em Tools and Algorithms for the Construction and Analysis of
  Systems - 22nd International Conference, {TACAS} 2016, Held as Part of the
  European Joint Conferences on Theory and Practice of Software, {ETAPS} 2016,
  Eindhoven, The Netherlands, April 2-8, 2016, Proceedings}, volume 9636 of
  {\em Lecture Notes in Computer Science}. Springer, 2016.

\bibitem{DBLP:conf/tacas/ChristakisL0W16}
M.~Christakis, K.~R.~M. Leino, P.~M{\"{u}}ller, and V.~W{\"{u}}stholz.
\newblock Integrated environment for diagnosing verification errors.
\newblock In Chechik and Raskin \cite{DBLP:conf/tacas/2016}, pages 424--441.

\bibitem{christakis2012collaborative}
M.~Christakis, P.~M{\"u}ller, and V.~W{\"u}stholz.
\newblock Collaborative verification and testing with explicit assumptions.
\newblock In {\em FM 2012: Formal Methods}, pages 132--146. Springer, 2012.

\bibitem{DBLP:conf/icse/Christakis0W16}
M.~Christakis, P.~M{\"{u}}ller, and V.~W{\"{u}}stholz.
\newblock Guiding dynamic symbolic execution toward unverified program
  executions.
\newblock In L.~K. Dillon, W.~Visser, and L.~Williams, editors, {\em
  Proceedings of the 38th International Conference on Software Engineering,
  {ICSE} 2016, Austin, TX, USA, May 14-22, 2016}, pages 144--155. {ACM}, 2016.

\bibitem{clarke2003counterexample}
E.~Clarke, O.~Grumberg, S.~Jha, Y.~Lu, and H.~Veith.
\newblock Counterexample-guided abstraction refinement for symbolic model
  checking.
\newblock {\em Journal of the ACM (JACM)}, 50(5):752--794, 2003.

\bibitem{czech2015just}
M.~Czech, M.-C. Jakobs, and H.~Wehrheim.
\newblock Just test what you cannot verify!
\newblock In {\em Fundamental Approaches to Software Engineering}, pages
  100--114. Springer, 2015.

\bibitem{d2010alloy+}
N.~D'Ippolito, M.~F. Frias, J.~P. Galeotti, E.~Lanzarotti, and S.~Mera.
\newblock Alloy+ hotcore: A fast approximation to unsat core.
\newblock In {\em Abstract State Machines, Alloy, B and Z}, pages 160--173.
  Springer, 2010.

\bibitem{filieri2013reliability}
A.~Filieri, C.~S. P{\u{a}}s{\u{a}}reanu, and W.~Visser.
\newblock Reliability analysis in symbolic pathfinder.
\newblock In {\em Proceedings of the 2013 International Conference on Software
  Engineering}, pages 622--631. IEEE Press, 2013.

\bibitem{henzinger2002lazy}
T.~A. Henzinger, R.~Jhala, R.~Majumdar, and G.~Sutre.
\newblock Lazy abstraction.
\newblock {\em ACM SIGPLAN Notices}, 37(1):58--70, 2002.

\bibitem{jackson2012software}
D.~Jackson.
\newblock {\em Software Abstractions: logic, language, and analysis}.
\newblock MIT press, 2012.

\bibitem{jhala2009software}
R.~Jhala and R.~Majumdar.
\newblock Software model checking.
\newblock {\em ACM Computing Surveys (CSUR)}, 41(4):21, 2009.

\bibitem{lal2014powering}
A.~Lal and S.~Qadeer.
\newblock Powering the static driver verifier using corral.
\newblock In {\em Proceedings of the 22nd ACM SIGSOFT International Symposium
  on Foundations of Software Engineering}, pages 202--212. ACM, 2014.

\bibitem{lal2012solver}
A.~Lal, S.~Qadeer, and S.~K. Lahiri.
\newblock A solver for reachability modulo theories.
\newblock In {\em International Conference on Computer Aided Verification},
  pages 427--443. Springer, 2012.

\bibitem{leino2010dafny}
K.~R.~M. Leino.
\newblock Dafny: An automatic program verifier for functional correctness.
\newblock In {\em Logic for Programming, Artificial Intelligence, and
  Reasoning}, pages 348--370. Springer, 2010.

\bibitem{pavese2010my}
E.~Pavese, V.~Braberman, and S.~Uchitel.
\newblock My model checker died!: how well did it do?
\newblock In {\em Proceedings of the 2010 ICSE Workshop on Quantitative
  Stochastic Models in the Verification and Design of Software Systems}, pages
  33--40. ACM, 2010.

\bibitem{tulsian2014mux}
V.~Tulsian, A.~Kanade, R.~Kumar, A.~Lal, and A.~V. Nori.
\newblock Mux: algorithm selection for software model checkers.
\newblock In {\em Proceedings of the 11th Working Conference on Mining Software
  Repositories}, pages 132--141. ACM, 2014.

\end{thebibliography}
\end{document}